\documentclass[%
  aps,
  prl,
  reprint,
  superscriptaddress,
  10pt,
  floatfix,
]{revtex4-2}

\usepackage{xcolor}
\usepackage{graphicx}
\usepackage{subcaption}
\usepackage{enumitem}
\usepackage{booktabs}
\usepackage{amsmath,amssymb}
\usepackage{import}
\usepackage{environ}
\usepackage{float}
\usepackage{placeins}
\usepackage{fvextra}
\usepackage{tikz}
\usepackage{pgfplots}
\pgfplotsset{compat=1.18}

\usepackage[colorlinks=true,linkcolor=blue!60!black,citecolor=blue!60!black,urlcolor=blue!60!black]{hyperref}
\usepackage[capitalize,noabbrev]{cleveref}

\newcommand{\KT}{KT1\xspace}

\usepackage{mfirstuc} % to uppercase the name
\usepackage{todonotes} % for the framed note boxes
\newcommand{\addReviewer}[2]{
  \expandafter\newcommand\csname #1\endcsname[1]{{\textbf{ \color{#2} \capitalisewords{#1}:\,##1}}}
  \expandafter\newcommand\csname #1cor\endcsname[2]{{\color{#2} \capitalisewords{#1}:\,\st{##1} \,{\textbf{##2}}}}
  \expandafter\newcommand\csname #1color\endcsname{#2}
  \expandafter\newcommand\csname #1todo\endcsname[1]{{\todo[inline,color=white!70!#2, caption={}]{\textbf{\capitalisewords{#1}}: ##1}}}
}

\usepackage{tikz, marginnote} % margin notes with circles
\newcommand{\checkedby}[1]{
\ifdefined\CROSSCHECKS
  \marginnote{
    \begin{tikzpicture}
      \foreach \x [count=\xi] in {#1} {
         \node[shape=circle,inner sep=0mm,
         minimum size=2mm,
         fill=\csname \x color\endcsname] at (\xi*3mm,0) {};
       }
    \end{tikzpicture}
  }
\else
\fi
}

\usepackage{soul,color}
\usepackage{xspace}

\definecolor{chromeyellow}{rgb}{1.0, 0.65, 0.0}
\definecolor{DodgeBlue}{rgb}{0.118, 0.565,1.000}
\definecolor{asparagus}{rgb}{0.53, 0.66, 0.42}
\definecolor{cadmiumgreen}{rgb}{0.0, 0.42, 0.24}
\definecolor{blue(ryb)}{rgb}{0.01, 0.28, 1.0}
\definecolor{periwinkle}{RGB}{181, 146, 203}
\definecolor{turquoiseblue}{rgb}{0.02, 0.55, 0.55}
\definecolor{green1}{RGB}{50,205,50}
\definecolor{amethyst}{rgb}{0.6, 0.4, 0.8}
\definecolor{indianred}{RGB}{205,92,92}

\addReviewer{anna}{orange}
\addReviewer{alex}{indianred}
\addReviewer{misha}{DodgeBlue}
\addReviewer{marian}{amethyst}
\addReviewer{ilya}{periwinkle}
\addReviewer{dhruv}{cadmiumgreen}

\begin{document}
% Called from paper.tex inside \begin{document} (revtex front matter).

\title{AI in Particle Physics Education: Research Problems and Foundational Skills}

\author{M. Mikhasenko}
\affiliation{Faculty of Physics and Astronomy, Ruhr-Universit\"at Bochum, D-44780 Bochum, Germany}
\author{M. Stahl}
\affiliation{Faculty of Physics and Astronomy, Ruhr-Universit\"at Bochum, D-44780 Bochum, Germany}
\author{I. Segal}
\affiliation{Faculty of Physics and Astronomy, Ruhr-Universit\"at Bochum, D-44780 Bochum, Germany}
\author{D. Parmar}
\affiliation{Faculty of Physics and Astronomy, Ruhr-Universit\"at Bochum, D-44780 Bochum, Germany}
\author{A. Zimmer}
\affiliation{Faculty of Physics and Astronomy, Ruhr-Universit\"at Bochum, D-44780 Bochum, Germany}
\author{A. Kazatsky}
\affiliation{Faculty of Physics and Astronomy, Ruhr-Universit\"at Bochum, D-44780 Bochum, Germany}

\date{\today}

\begin{abstract}
  Generative AI can produce perfect solutions to standard physics homework,
  weakening the connection between submitted work and knowledge a student can
  use independently. We report a redesign of the introductory nuclear and
  particle physics course at Ruhr University Bochum, in which AI was permitted
  and encouraged alongside traditional tools on unusually difficult,
  research-shaped assignments. We found that most students remained engaged
  with the assignments throughout the course and produced ambitious work,
  while the course did not reliably secure its foundational objective.
  We offer this mixed experience as a reference point for discussion.
\end{abstract}

\maketitle

\section{Introduction}
\label{sec:introduction}

Generative AI has undermined a basic assumption of physics education: a submitted assignment solution no longer establishes a student's capability to solve the problems independently.
A coherent derivation and solution can now be produced without the student working through the reasoning that the exercise was meant to develop.
When producing an answer becomes separate from understanding it,
instructors must reconsider what homework teaches and
what a completed solution reveals about the student's abilities~\cite{Kortemeyer2023,Lodge2023}.
AI-based tools were already widely used by university students when this course
was taught~\cite{vonGarrel2023}.
Physics education therefore faces two risks at once:
\textit{(i)}~allowing students to rely on AI for a single-prompt answer without
gaining the intended knowledge and
\textit{(ii)}~failing to teach students how
to use and question AI-generated answers.

The introductory nuclear and particle physics course, hereafter referred to as KT1, at
Ruhr University Bochum explored the integration of AI in the winter semester 2025/26.
Ten homework assignments of unusual difficulty were given to students, who were encouraged to use AI alongside books, discussion, and staff support.
Instead of routine applications of one known method, the assignments were
designed as undergraduate-level research problems that combined several
methods, data representations, and workflows.

We designed the course around the instructor's experience that much scientific
knowledge is acquired while working on open-ended problems of higher complexity, see also~\cite{Ogilvie:2009,Brookes:2020}.
We asked whether AI could allow students to enter this learning process earlier: not simply to obtain an answer, but to question, test, correct,
and make sense of the output.
We sought to engage and challenge students, requiring their active involvement.
This approach offered an alternative to evaluating perfectly written solutions to standard problems.

\KT is an elective 6\,CP course with
four lecture hours and two tutorial hours per week.
For the presentation of the material we decided to use a top-down narrative
that began with the modern picture and then moved across sectors of the Standard Model one by one.
The textbook "Modern Particle Physics" by M.~Thomson~\cite{Thomson:2013zua} served as a reference.
The participant cohort was significant: forty-two students handed in at least one exercise sheet, 22 of whom were in the fifth semester of their bachelor's studies.

The course design separated three types of activity.
\textit{Lectures} introduced concepts, notation, and physical motivation.
Every lecture started with short recap questions that drew from previously covered material
and combined conceptual and computational problems.
\textit{Tutorials} were intended to provide practice in solving standard analytic problems.
The problems were disclosed after the lecture, and students were asked to solve
them in class, with each student taking a turn to write a solution on the
blackboard during the tutorials.
\textit{Homework} was the exploratory component.
The tasks asked for working with literature, modeling, plotting,
detector interpretation, derivations, analysis-style measurements, and
scientific communication.
Students were given 1.5 weeks to complete each assignment.
AI was explicitly allowed and introduced in the first lecture as a potential
partner and teacher, with students remaining responsible for verifying the result.
No licenses were provided by the course,
but students had access to several models with mid-range capabilities:
free models with strict token limitations (ChatGPT, Claude),
as well as GPT@RUB~\cite{GPTRUB} and Gemini 3 Pro with more generous usage limits.
For each sheet, the best submission selected by the assistants received bonus points.
The \textit{written examination} was individual, AI- and tools-free, with 90 minutes for eight problems.
Two paths for successful evaluation of the course were offered to students:
a homework- or examination-heavy weighting, 70/30 in either direction.
At the evaluation stage, the better result from the two schemes was retained.
Ultimately, all students who passed the course were graded under the homework-heavy scheme.

This report presents and evaluates four local sources produced at the time of the course.
Course documents establish the assignment and assessment design.
The administrative record provides homework, bonus points, examination,
and final outcome data.
A mid-semester survey was carried out to gauge students' perceptions of the new format.
Finally, instructor and staff notes record observations and the reasons for
changes made during and after the semester.

While reflecting on the course experience and preparing this report, we found a large corpus of literature relating to physics education, the integration of research into teaching, and the role of AI in education. We cite relevant work alongside the aspects of the course that it helped us interpret and to support readers preparing comparable courses.

\section{Undergraduate-level research problems}
\label{sec:problems}
The assignments brought selected elements of research practice into self-contained undergraduate problems.
Professional research may take months,
whereas the assignments in this course were
self-contained pieces of that process, thus are referred to as \emph{research-shaped}.
Such an approach has shown promising effects on students' academic and professional development~\cite{Linn2015,Auchincloss2014}.

The assignments deliberately covered a wide range of subjects and types of work.
Some were focused on one facet of research,
while others required a broader range of skills
and had the potential to develop into junior research projects.
Table~\ref{tab:hw-main} shows the full sequence,
and the complete problem formulations are given in Appendix~\ref{app:homework}.
\newcommand{\hwfocus}[1]{\parbox[t]{0.46\textwidth}{\raggedright #1}}
\begin{table*}[t]
  \centering
  \caption{Ten homework sheets issued in WS~25/26. All sheets are shown in the Appendix~\ref{app:homework}.}
  \label{tab:hw-main}
  \small
  \begin{tabular}{clll}
    \toprule
    HW & Topic             & Group size & Task focus \\
    \midrule
    1  & PDG numbering     & $\leq 3$   & \hwfocus{Particle numbering and quantum-number logic} \\
    2  & $pp$ flux / 3D    & $\leq 2$   & \hwfocus{Angular flux model and spherical visualisation} \\
    3  & Belle II event    & $\leq 4$   & \hwfocus{Event display and detector interpretation} \\
    4  & Calorimeter       & $\leq 4$   & \hwfocus{Detector acceptance, resolution, and design constraints} \\
    5  & Dirac spinors     & $\leq 3$   & \hwfocus{Physical parameters of spinor states} \\
    6  & Rosenbluth        & individual & \hwfocus{Complete analytic derivation} \\
    7  & Flavour SU(3)     & $\leq 3$   & \hwfocus{Symmetry relation and hadron phenomenology} \\
    8  & Hadron characters & unlimited  & \hwfocus{Conceptual visual system for excited hadrons} \\
    9  & $W$ mass/width    & $\leq 3$   & \hwfocus{Measurement-style analysis task} \\
    10 & Nuclear podcast   & $\leq 6$   & \hwfocus{Nuclear physics communication task} \\
    \bottomrule
  \end{tabular}
\end{table*}
We discuss a selected set of assignments
to give the reader an idea of the work.

\begin{figure}[t]
  \centering
  \includegraphics[width=0.65\linewidth]{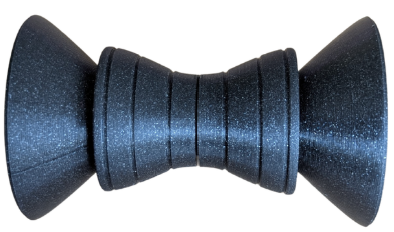}
  \caption{3D-printed model of the instructor's reference solution for the
    HW2 proton--proton charged-particle angular-flux calculation.
    Cylindrical coordinates with double log scale ($\log r,\,\phi,\,\log z$) are used to represent all features of the process. The grooves show fixed values of pseudorapidity.}
  \label{fig:hw2-object}
\end{figure}

HW2 represented the most research-like end of the problem set.
Students had to extract or estimate information from published
elastic and inelastic proton--proton collision measurements,
transform the reported variables to an angular flux,
determine the relative normalizations,
combine forward elastic peaks with the broader inelastic component, and make a
spherical surface plot.
The route was only partly specified and involved multiple challenges,
including choosing common variables (the inclusive flux is given differentially in rapidity, whereas the exclusive reaction is parametrized by the Mandelstam variable $t$) and reconciling datasets that covered different ranges.
Visual representation was also difficult,
as the quantities differ by several orders of magnitude: elastic scattering produces a significant probability density concentrated at polar angles below $1\,$mrad.
The solution, shown in Fig.~\ref{fig:hw2-object}, was 3D-printed using institutional facilities.

HW4 was close to a conventional calculation, but at the same time a typical research task.
Given photon four-vectors from simulated $B^+\to\pi^0\pi^+$ and
$D^0\to K^+K^-\pi^-\pi^+\pi^0$ decays, students estimated calorimeter cell
size, simulated energy smearing, reconstructed the $\pi^0\to\gamma\gamma$ mass peak, and considered engineering constraints.
The physics route was bounded, but it required four-vector handling, detector
geometry, histograms, random sampling, interpretation and literature review.

HW6 provided a contrasting individual analytic task: the full Rosenbluth
derivation for elastic electron--proton scattering with Dirac and Pauli form
factors.
It could be completed through a long trace calculation or a shorter argument
based on the allowed tensor structures.
The task required sustained symbolic control and explicit kinematic reasoning.

\begin{figure}[b]
  \centering
  \includegraphics[width=0.95\linewidth]{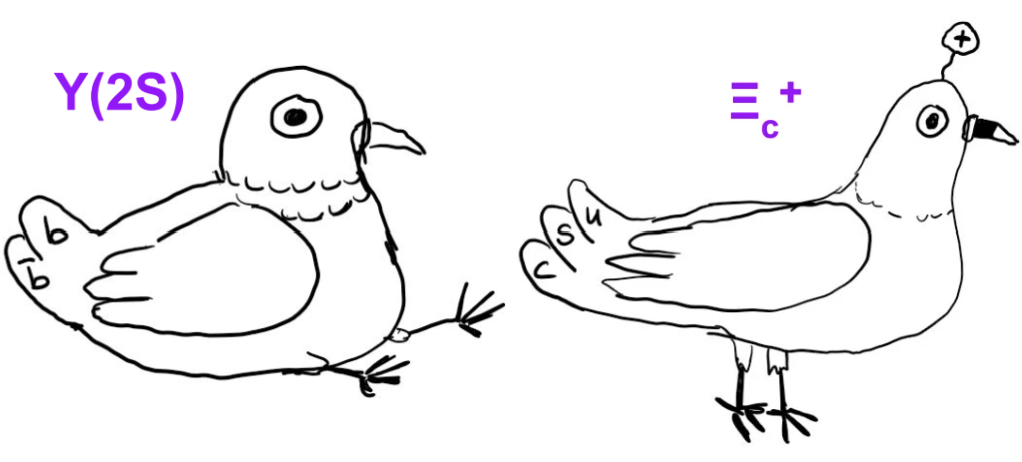}
  \caption{
  Example character pair prepared for HW8 by teaching assistant A. Zimmer.
  Standing pigeons with three tail feathers are baryons, and
  sitting pigeons with two tail feathers are mesons. Each feather contains one of the quarks. The number of neck ruffs decodes radial excitations,
  and stripes on wings indicate orbital excitations. The charge is depicted by baubles on the head, while isospin can be seen by the coloring of the beak. A white beak refers to $I=0$, a half black and half white beak is $I=\frac{1}{2}$ and a fully black beak denotes $I=1$.
  }
  \label{fig:hadron-characters}
\end{figure}

HW8 called for the student's creativity.
Over the Christmas break, students were asked to design consistent visual characters
representing excited hadrons,
translating flavor content, spin and parity, radial and orbital excitation, thresholds,
decays, and experimental history into character features.
The exercise was deliberately playful while remaining academic at its core,
closer to an outreach task, which is an important aspect of fundamental research.
Figure~\ref{fig:hadron-characters} shows one example of the resulting excited-hadron creatures.

\subsection{The intended role of AI}
The use of generative AI was encouraged to explain a concept, propose a decomposition,
generate or debug code, manipulate algebra, and help interpret an error beyond the classical resources of books, scientific papers and tutorials.
HW4 and HW10 explicitly required citations,
giving students the opportunity to discover that models tend to make them up~
\footnote{
  For modern tool-augmented LLM systems on summer 2026,
  fabricated references are no longer a practical limitation~\cite{Asai:2026,Rao:2026}.
}.

The intended student responsibility was to decide whether the result made physical sense:
for a PDG-numbering program (HW1), students had to test their programs for correct quantum numbers and exceptions;
an angular distribution (HW2) had to capture the correct relative probabilities of elastic and inelastic scattering;
an event-display interpretation had to be consistent with tracks, clusters, vertices, and conservation laws.

The assignments were also a response to the capabilities of models available in late 2025.
The instructor tested candidate tasks before distribution and found that a single prompt did not reliably yield a complete solution.
No systematic benchmark of named models, prompts, or access tiers was conducted.
The paper treats resistance to a single prompt as design experience, not as a durable property of the assignments.

As models have improved drastically since then, the educational value must lie in the reasoning and checking required
by the task rather than in a continuous race to remain unpromptable.
At the time of publication, Codex 5.6-Sol can produce solutions to HW1, HW4, HW5, HW6, HW7, and HW8.
It struggles with HW2 (a complex multistep workflow), HW3 (a visual interface), HW9 (human input for visualization is needed), and HW10 (no multi-speaker podcast capability).

\section{Development and reception of the course}
\label{sec:evaluation}

Student engagement with the course was visible in forms that the grades alone do not capture.
Several HW1 groups built reusable libraries or command-line tools, going far beyond the work required for full credit.
One HW6 derivation exceeded 15 pages and extended more than two meters when displayed across the blackboards;
and several HW10 groups produced artistically ambitious scientific podcasts.
Such engagement was not confined to a few exceptional cases.
Out of 42 active coursework records,
33 received positive credit on at least eight
of the ten sheets and 24 received credit on all ten.

A questionnaire after the first three assignments had 30 complete responses.
All described the homework as hard or very hard, and
13 reported spending at least eight hours per week on it.
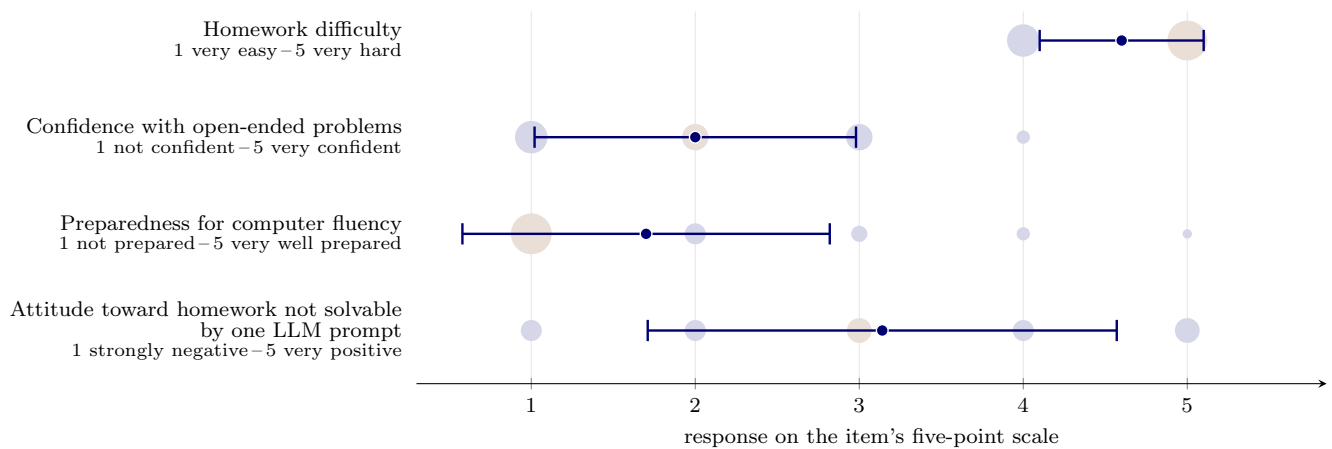
\begin{figure*}[t]
  \centering
  \begin{tikzpicture}
    \begin{axis}[
      width=0.76\textwidth, height=6.5cm,
      xmin=0.30, xmax=5.85,
      xtick={1,2,3,4,5},
      ymin=0.45, ymax=4.30,
      ytick={1,2,3,4},
      yticklabels={%
        {Attitude toward homework not solvable\\[-0.2em]by one LLM prompt\\[-0.25em]{\scriptsize 1 strongly negative\,--\,5 very positive}},%
      {Preparedness for computer fluency\\[-0.25em]{\scriptsize 1 not prepared\,--\,5 very well prepared}},%
      {Confidence with open-ended problems\\[-0.25em]{\scriptsize 1 not confident\,--\,5 very confident}},%
      {Homework difficulty\\[-0.25em]{\scriptsize 1 very easy\,--\,5 very hard}}},
      yticklabel style={align=right,font=\footnotesize},
      xlabel={response on the item's five-point scale},
      xlabel style={font=\footnotesize},
      tick label style={font=\footnotesize},
      axis lines=left,
      y axis line style={draw=none},
      ytick style={draw=none},
      xmajorgrids, grid style={draw=black!10},
      clip=false,
      ]

      % Response distributions: disc area proportional to the number of respondents.
      % The median disc in each row is light orange.
      \fill[blue!45!black!16] (axis cs:4,4) circle [radius=0.215cm];
      \fill[orange!45!black!16] (axis cs:5,4) circle [radius=0.263cm];
      \fill[blue!45!black!16] (axis cs:1,3) circle [radius=0.215cm];
      \fill[orange!45!black!16] (axis cs:2,3) circle [radius=0.175cm];
      \fill[blue!45!black!16] (axis cs:3,3) circle [radius=0.175cm];
      \fill[blue!45!black!16] (axis cs:4,3) circle [radius=0.088cm];
      \fill[orange!45!black!16] (axis cs:1,2) circle [radius=0.270cm];
      \fill[blue!45!black!16] (axis cs:2,2) circle [radius=0.139cm];
      \fill[blue!45!black!16] (axis cs:3,2) circle [radius=0.107cm];
      \fill[blue!45!black!16] (axis cs:4,2) circle [radius=0.088cm];
      \fill[blue!45!black!16] (axis cs:5,2) circle [radius=0.062cm];
      \fill[blue!45!black!16] (axis cs:1,1) circle [radius=0.139cm];
      \fill[blue!45!black!16] (axis cs:2,1) circle [radius=0.139cm];
      \fill[orange!45!black!16] (axis cs:3,1) circle [radius=0.164cm];
      \fill[blue!45!black!16] (axis cs:4,1) circle [radius=0.139cm];
      \fill[blue!45!black!16] (axis cs:5,1) circle [radius=0.164cm];

      % Mean with one standard deviation.
      \draw[blue!45!black,line width=0.9pt] (axis cs:4.10,4) -- (axis cs:5.10,4);
      \draw[blue!45!black,line width=0.9pt] (axis cs:4.10,3.89) -- (axis cs:4.10,4.11);
      \draw[blue!45!black,line width=0.9pt] (axis cs:5.10,3.89) -- (axis cs:5.10,4.11);
      \draw[blue!45!black,line width=0.9pt] (axis cs:1.02,3) -- (axis cs:2.98,3);
      \draw[blue!45!black,line width=0.9pt] (axis cs:1.02,2.89) -- (axis cs:1.02,3.11);
      \draw[blue!45!black,line width=0.9pt] (axis cs:2.98,2.89) -- (axis cs:2.98,3.11);
      \draw[blue!45!black,line width=0.9pt] (axis cs:0.58,2) -- (axis cs:2.82,2);
      \draw[blue!45!black,line width=0.9pt] (axis cs:0.58,1.89) -- (axis cs:0.58,2.11);
      \draw[blue!45!black,line width=0.9pt] (axis cs:2.82,1.89) -- (axis cs:2.82,2.11);
      \draw[blue!45!black,line width=0.9pt] (axis cs:1.71,1) -- (axis cs:4.57,1);
      \draw[blue!45!black,line width=0.9pt] (axis cs:1.71,0.89) -- (axis cs:1.71,1.11);
      \draw[blue!45!black,line width=0.9pt] (axis cs:4.57,0.89) -- (axis cs:4.57,1.11);

      \filldraw[fill=blue!45!black,draw=white,line width=0.5pt] (axis cs:4.60,4) circle [radius=2.2pt];
      \filldraw[fill=blue!45!black,draw=white,line width=0.5pt] (axis cs:2.00,3) circle [radius=2.2pt];
      \filldraw[fill=blue!45!black,draw=white,line width=0.5pt] (axis cs:1.70,2) circle [radius=2.2pt];
      \filldraw[fill=blue!45!black,draw=white,line width=0.5pt] (axis cs:3.14,1) circle [radius=2.2pt];
    \end{axis}
  \end{tikzpicture}
  \caption{Four mid-semester items
    ($n=30$ for the upper three items, $n=29$ for the lowest). Discs show the response distribution with area
    % both size, and scale would be incorrect
    % it is area ~ scale^2
    proportional to the number of respondents; the bar gives the mean and one
    standard deviation, and the light-orange disc the median. Because the responses
    are bounded ordinal ratings, the mean and standard deviation are descriptive
    only. The items measure perceptions, not demonstrated competence, and the
    eligible denominator at the survey date is unknown.}
  \label{fig:survey-likert}
\end{figure*}

Reported LLM use was frequent:
24 of 29 respondents selected often or always, and none selected never.
Attitudes toward assignments designed not to be solvable by a single LLM prompt were divided:
12 responses were positive, ten negative, and seven neutral.
Figure~\ref{fig:survey-likert} also shows low self-reported preparedness for the
required computer fluency and low confidence with open-ended problems.
These are important distinctions: perceived learning and confidence can diverge
from measured learning even in introductory physics~\cite{Sitzmann2010,Deslauriers2019}.
The survey does not indicate that LLM use increased confidence.

\subsection{Where the design created friction}

The first difficulty appeared with HW1.
Its particle-identification problem was intended to reveal the organization of
the particle spectrum by turning the PDG numbering rules into a program.
Many students instead encountered programming as an undeclared prerequisite,
consistent with the survey responses in Fig.~\ref{fig:survey-likert}.
In response, we reduced some later tooling demands,
but retained the central idea of research-shaped homework.

Time was a second problem. Research-shaped work needs time not only to obtain an output but to understand
why it is correct.
Respondents asked for longer deadlines, clearer deliverables, stronger links to
the lectures, and dedicated help.
When a student lacked sufficient physics knowledge to evaluate the
generated answer,
the intended conversation with AI could become a cycle of guessing in response to plausible but unreliable answers.
This is an instructor interpretation supported by the reported difficulty and requests for support.
Open responses also raised concerns about dependence on AI and unequal access to paid models.
At the same time, the complexity of the assignments increased the teaching assistants' workload.

The tutorials also created friction. Disclosing the problems in advance gave
students time to prepare AI-generated solutions, which some then copied onto
the board during the tutorials without engaging in the intended reasoning
process. In addition, staff notes and survey themes
indicate that many students found taking their turn at the blackboard
intimidating.
Open survey responses requested more preparation time, more stable group structures, and stronger alignment among lectures, tutorials, homework, and examination.

The examination scores were not obtained under neutral inputs.
For students with substantial homework and bonus credit,
the score formula reduced the marginal effect of maximizing an exam score.
This incentive structure prevents the examination from serving as a clean learning measure or an estimate of an AI effect.
Moreover, homework was commonly collaborative and resource-rich,
whereas the exam had a conventional format.
Furthermore, previous iterations of the course did not require a written examination. To compensate, the tutors released two mock examinations.
These factors do not make the examination record irrelevant.
The 27 students who took the examination received a mean score of 20.6 out of 80,
and only two reached 40 points. Many students left problems unattempted because they knew their homework scores had already secured passing the threshold; this behavior skewed the mean score.
The instructor also observed that some students who needed examination points
and made a serious attempt were unable to complete standard calculations.
That observation was not collected through a formal protocol, but it made the
low record difficult to dismiss as incentive alone.
The examination therefore became a serious warning about students' unaided performance.

\section{Discussion and Conclusion}
\label{sec:discussion}
Students in this introductory course sustained work on undergraduate-level research problems and produced ambitious visualizations, extended derivations, high-quality reports, and podcasts among other results.
However, the course did not establish that they could reproduce foundational calculations without assistance; the examination provided a serious but confounded warning.

This experience reinforced a distinction well established in the education literature: assisted performance is not the same as independently retrievable knowledge; therefore, AI may act as an effective tutor~\cite{VanLehn2011}, but students need dedicated unaided practice to obtain foundational knowledge.
The prevalent educational criterion needs to be whether a student can recall and apply knowledge without aid, as is the case in a written examination~\cite{Soderstrom2015}.
Experimental evidence outside this course likewise shows that unrestricted access to answers can decrease long-term learning outcomes~\cite{Bastani2025}.
In a physics-specific study, students using unrestricted ChatGPT frequently accepted incorrect answers and relied heavily on direct copy-and-paste queries~\cite{Krupp2024}.
By contrast, a purpose-built and scaffolded AI tutor improved immediate learning relative to an in-class active-learning lesson~\cite{Kestin2025}.

Since the accuracy of LLMs will continue to improve, we believe that AI literacy should be taught to students early, as supported by other investigations~\cite{Kasneci2023}. This includes how to verify generated answers and how to use AI as a teacher instead of blindly copying its output.
Furthermore, the course experience showed that exposure to research problems and foundational knowledge are distinct educational achievements and cannot be assumed to train or demonstrate one another.

The present plan for the next \KT course reflects our observations.
\textit{(i)}~A written examination will be the focus of the course and will determine the final grade. Homework will count only toward bonus points. To support student learning, multiple mock examinations will be offered so that students can practice solving standard problems without assistance~\cite{RoedigerKarpicke2006}.
\textit{(ii)}~Traditional tutorial problems will be given to students well in advance so that they can come prepared and participate in active classroom discussions. This shifts the focus from encouraging independent work toward the group discussion made possible by preparation~\cite{HellerKeithAnderson1992,HellerHollabaugh1992}.
\textit{(iii)}~Parts of the research-shaped problems may remain as advanced or bonus work, and AI will remain available during study. These tasks will be accompanied by prerequisite preparation, worked examples, feedback, and explicit consolidation~\cite{Kirschner2006,Alfieri2011}.

Beyond these immediate changes, the experience raises three broader questions for the physics community.
\textit{(i)}~What foundational knowledge should future physicists be able to reproduce unaided?
\textit{(ii)}~To what extent should research-shaped tasks be part of an undergraduate course?
\textit{(iii)}~How should physics education cultivate commonly required scientific skills?

% \begin{acknowledgments}
\textit{Acknowledgments.}
  We thank the students whose sustained participation and course feedback made
  this reflection possible.
  % We also acknowledge 
  % the teaching assistants and correctors whose support and evaluation work were integral to the course.
% \end{acknowledgments}

\textit{Use of AI.}
This paper was prepared with assistance from Codex GPT-5.5 and Cursor 2.5 for source inspection, quantitative checks, literature discovery, speech to text conversion, and language revision. The authors verified the analysis and references and retain responsibility for the paper.

\bibliographystyle{apsrev4-2}
\bibliography{references}

\onecolumngrid
\setcounter{secnumdepth}{4}
\appendix

\section{Homework assignments issued in WS~25/26}
\label{app:homework}

The exercise sheets are attached on the following pages exactly as they were issued
to students.

% Each issued page is placed as a fitted graphic rather than with \includepdf:
% revtex's reprint mode is two-column, so pdfpages ships each PDF page as a
% column and stacks two per sheet of paper. Going through \includegraphics
% keeps normal page building, one issued page per printed page.
% The issued files are letter-size with ~1in margins; those margins are cropped
% so the content, not the original paper, fills the appendix page.
\newcommand{\sheetfilename}[1]{sheets/sheet_\ifnum#1<10 0\fi#1.pdf}
% #1 homework number, #2 page in that sheet, #3 pages in the sheet.
% Letter-class margins are cropped (ink bbox ~71/42/71/48 bp on 612x792).
\newcommand{\sheetpage}[3]{%
  \clearpage
  \thispagestyle{plain}%
  \begingroup
    \setlength{\fboxrule}{0.5pt}%
    \setlength{\fboxsep}{10pt}%
    \centering
    \fcolorbox{black!20}{white}{%
      \includegraphics[%
        page=#2,%
        trim=68bp 55bp 68bp 48bp,%
        clip,%
        width=\dimexpr\textwidth-2\fboxsep-2\fboxrule\relax,
        height=\dimexpr\textheight-3.2em-2\fboxsep-2\fboxrule\relax,
        keepaspectratio
      ]{\sheetfilename{#1}}}%
    \par\vspace{0.7em}%
    {\footnotesize\color{black!50}%
      HW~#1%
      \ifnum#3>1
        \quad page~#2 of~#3%
      \fi}%
    \par
  \endgroup
}

% Homework 1: Particle numbering scheme
\sheetpage{1}{1}{1}

% Homework 2: Differential cross section in different variables
\sheetpage{2}{1}{2}
\sheetpage{2}{2}{2}

% Homework 3: Reconstruct an event in a detector
\sheetpage{3}{1}{2}
\sheetpage{3}{2}{2}

% Homework 4: Calorimeter properties
\sheetpage{4}{1}{1}

% Homework 5: Physical parameters of a Dirac spinor
\sheetpage{5}{1}{2}
\sheetpage{5}{2}{2}

% Homework 6: Derivation of the Rosenbluth formula
\sheetpage{6}{1}{2}
\sheetpage{6}{2}{2}

% Homework 7: Flavor SU(3) relation
\sheetpage{7}{1}{3}
\sheetpage{7}{2}{3}
\sheetpage{7}{3}{3}

% Homework 8: Hadron characters outreach
\sheetpage{8}{1}{4}
\sheetpage{8}{2}{4}
\sheetpage{8}{3}{4}
\sheetpage{8}{4}{4}

% Homework 9: Measurements of $W$ mass and width
\sheetpage{9}{1}{1}

% Homework 10: Nuclear podcast
\sheetpage{10}{1}{3}
\sheetpage{10}{2}{3}
\sheetpage{10}{3}{3}

\end{document}